# Geometric phase metasurfaces for linearly polarized light

Yubin Gao[1] (3190101611@zju.edu.cn), Qikai Chen[1] (12330005@zju.edu.cn),

Yaoguang Ma[1,*] (mayaoguang@zju.edu.cn)

## Affiliations

[1] State Key Laboratory for Extreme Photonics and Instrumentation, College of Optical Science and Engineering, ZJU–Hangzhou Global Scientific and Technological Innovation Center, Intelligent Optics and Photonics Research Center, Jiaxing Research Institute, Zhejiang University, Hangzhou 310027, China

[*] Corresponding author. Email: mayaoguang@zju.edu.cn. Tel: +86-18626859918

# Abstract

The geometric phase is a universal concept in modern physics and has enabled the development of metasurfaces for versatile wavefront shaping. However, its realization in metasurfaces has been restricted to circularly polarized light, confining geometric phase metasurfaces to helicity-dependent operation and excluding them from the linear-polarization domain that dominates modern optics. In this work, we overcome this limitation by harnessing exceptional points of non-Hermitian physics. We introduce and experimentally realize quasi-exceptional-point metasurfaces that exploit engineered singularities to directly impart a geometric phase onto linearly polarized light. Proof-of-principle demonstrations with gratings and holograms confirm broadband and high-fidelity wavefront shaping across arbitrary linear polarizations, which has not been achieved with previous phase modulation approaches. By revealing an intrinsic connection between geometric phase and non-Hermitian photonics, our work resolves a long-standing theoretical impasse and establishes a new framework for high-dimensional light control, opening opportunities for scalable polarization optics, advanced imaging, holography, optical communications, and integrated photonics.

# Introduction

The geometric phase, a ubiquitous phenomenon woven into the fabric of wave physics, arises not from dynamical propagation but from the innate geometry of a system's parameter space, manifesting when a wave undergoes a slow, cyclic evolution upon its rapid oscillation[1]. From its early intimations in the Fresnel-Arago laws of interference in 1819[2] to its seminal formalization by Berry in quantum mechanics in 1984[3], the geometric phase has transcended its original context to emerge as a universal paradigm throughout the physical sciences, underpinning phenomena from quantum entanglement to classical optics[4–6].

In optics, the geometric phase manifests elegantly when the polarization state traces a closed path on the Poincaré sphere, yielding a phase shift equal to half the enclosed solid angle[7,8]—a profound manifestation known as the Pancharatnam–Berry (PB) phase. Leveraging this principle, geometric phase metasurfaces (GPMs) have emerged as a compact and flexible platform for wavefront engineering[9]. By precisely sculpting subwavelength structures that impart spatially variant PB phases, GPMs have enabled a broad spectrum of applications, including advanced imaging[10–12], holography[13,14], structured light generation[15,16], quantum optics[17,18], and ultra-precise metrology[19,20].

Yet, beneath this veneer of progress—including the recent development of high-order GPMs[21,22]—lies a profound and persistent limitation that has continued to stifle the field's evolution: the conventional PB phase is intrinsically helicity dependent, restricting the operation of GPMs exclusively to circularly polarized (CP) light[23]. This fundamental impediment has severely curtailed the versatility of GPMs and, most critically, erected a formidable barrier to their integration into the linearly polarized (LP) photonic platforms that underpin mainstream optics. This chasm between a revolutionary technology and the established LP platforms has remained a central challenge, leaving a vast potential untapped.

Meanwhile, advances in non-Hermitian photonics have unveiled the unique role of exceptional points (EPs). These singularities, where eigenvalues and corresponding eigenstates coalesce simultaneously in parameter space, endow optical systems with an unprecedented ability to manipulate their eigenmodes[24,25]. Metasurfaces, as inherently open systems that exchange energy with their environment, provide a natural and fertile ground for embedding such singularities through judiciously tailored resonance modes[26], scattering matrices[27], or Jones matrices[28,29].

Here, we shatter the persistent helicity constraint of the PB phase by forging an alliance between the two seemingly disparate worlds of geometric phase and non-Hermitian physics. We report, for the first time, a universal strategy to engineer and imprint geometric phases directly onto LP light by incorporating EP-driven eigen-polarization control into metasurface design. Our approach leverages the inherent anholonomy of polarization evolution around engineered EPs within quasi-EP meta-atoms, enabling us to transcend the traditional circular polarization requirement. We experimentally validate this concept through demonstrations of GPM gratings and holograms operating under linear polarization. This work establishes a new paradigm to address a long-standing divide and unlocks a new era of versatile, broadband, and efficient polarization and wavefront control for the next generation of advanced photonic systems.

## Results

### Working principle and numerical simulations

The eigen-polarizations of a meta-atom—orthogonal eigenstates of its Jones matrix—determine how incident light couples to, interacts with, and evolves through the structure. Conventional GPMs are confined to circular polarization because their operation relies on orthogonal linear polarizations as the eigenstates of the meta-atom.

As illustrated in Fig. 1a, such metasurfaces impart a geometric phase to the output polarization component with handedness opposite to the incident CP wave,

$$\phi_{\text{conventional GPM}} = 2\sigma\theta, \quad (1)$$

where $\sigma = \pm 1$ denotes left- and right-handed circular polarization (LCP and RCP), respectively, and $\theta$ is the in-plane rotation angle of the periodically arranged meta-atoms (Supplementary Note 1). To realize Eq. (1), eigen-polarizations of the rotated meta-atom must form two orthogonal linear polarizations oriented at an angle $\theta$ (Fig. 1b). In this case, the polarization evolution from incident LCP to output RCP (or vice versa for RCP input) for two meta-atoms rotated by 0 and $\theta$ traces a closed path on the Poincaré sphere. The enclosed solid angle is $4\sigma\theta$, and half of this value is equal to the geometric phase in Eq. (1).

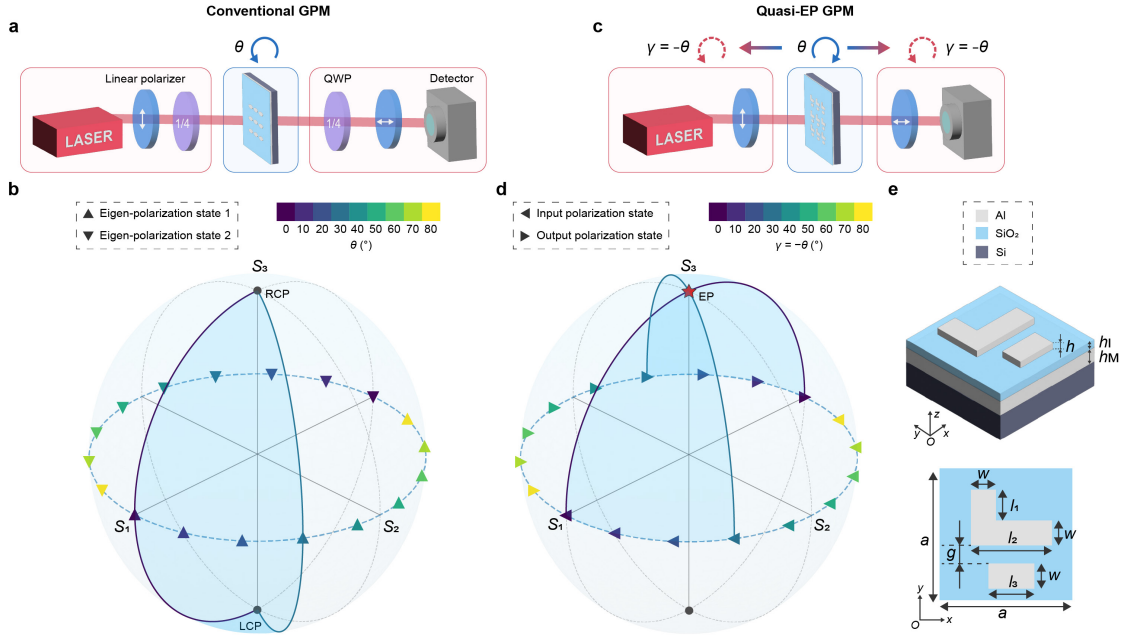

**Fig. 1 | Principle of conventional GPMs and quasi-EP GPMs. a,** Schematic of the experimental configuration for conventional GPMs. QWP, quarter-wave plate. **b,** Polarization evolution on the

Poincaré sphere from incident LCP (RCP) to output RCP (LCP) for meta-atoms rotated by 0 and $\theta$ (30° in this example). The purple and blue solid curves indicate the evolution trajectories, and the enclosed area (blue) corresponds to the geometric phase in Eq. (1). Upward- and downward-pointing triangular markers denote the two orthogonal linear polarization eigenstates of meta-atoms at different rotation angles. **c,** Schematic of the experimental configuration for quasi-EP GPMs. When a meta-atom is rotated by an angle $\theta$, the relative rotation can be equivalently described as fixing the meta-atom while both the incident and output polarization states rotate by $\gamma = -\theta$. **d,** Polarization evolution on the Poincaré sphere from LP inputs at 0 and $\gamma$ (30° in this example) to their orthogonal LP outputs, for a meta-atom tuned to an RCP EP. Trajectories (purple and blue solid curves) enclose a blue-shaded region with the same area as in (**b**), thereby producing an identical geometric phase. Leftward- and rightward-pointing triangular markers denote the linear polarization states of the incident and output waves at different orientations. **e,** Perspective (top) and top view (bottom) of a quasi-EP GPM unit cell. Geometric parameters: $h = 40$ nm, $h_\mathrm{I} = 50$ nm, $h_\mathrm{M} = 150$ nm, $a = 400$ nm, $l_1 = 100$ nm, $l_2 = 260$ nm, $l_3 = 145$ nm, $w = 80$ nm, and $g = 60$ nm.

We overcome this constraint by harnessing eigen-polarization degeneracies at EPs. For a single-layer metasurface, the out-of-plane symmetry dictates that the Jones matrix of a meta-atom, expressed in the linear polarization basis, can be written as

$$\mathbf{J} = \begin{bmatrix} A & C \\ C & B \end{bmatrix}, \tag{2}$$

where $A$, $B$, and $C$ are complex coefficients. When the condition

$$A - B + 2i\sigma C = 0, \tag{3}$$

is satisfied, the eigenvalues and eigenvectors of **J** coalesce, giving rise to an EP in polarization, where the degenerate eigenstate corresponds to LCP ($\sigma = +1$) or RCP ($\sigma = -1$) (Supplementary Note 2).

To illustrate, we consider a meta-atom at an RCP EP and analyze the output component that is orthogonal to a given LP input. As shown in Fig. 1c, because the rotation is inherently relative, rotating the meta-atom by an angle $\theta$ is equivalent to fixing the metasurface and rotating both the incident and output LP directions together by an angle $\gamma = -\theta$. In Fig. 1d, for LP inputs oriented at 0 and $\gamma$, the polarization evolution from the input state to its orthogonal counterpart is forced to pass through the RCP state due to the degeneracy of eigen-polarizations. The resulting trajectories (purple and blue solid curves) enclose a region on the Poincaré sphere identical in area to that in Fig. 1b, thereby yielding the same geometric phase.

A rigorous Jones-matrix analysis (Supplementary Note 2) shows that for an incident LP wave with Jones vector $[\cos\gamma \quad \sin\gamma]^\mathrm{T}$, the output after interaction with a rotated meta-atom is

$$\vec{O} = e^{i\sigma\tilde{\theta}}(A\cos\tilde{\theta} - i\sigma B\sin\tilde{\theta})\begin{bmatrix}\cos\gamma\\ \sin\gamma\end{bmatrix} + \frac{1}{2}i\sigma(A-B)e^{i2\sigma\tilde{\theta}}\begin{bmatrix}-\sin\gamma\\ \cos\gamma\end{bmatrix}, \quad (4)$$

where $\tilde{\theta} = \gamma - \theta$ is the relative rotation angle between the rotated linear-polarization reference frame (by $\gamma$) and the orientation of the meta-atom (by $\theta$). Equation (4) demonstrates that when the meta-atom is rotated by a relative angle $\tilde{\theta}$, the cross-polarized component acquires a geometric phase

$$\phi_\mathrm{quasi-EP\ GPM} = 2\sigma\tilde{\theta}. \quad (5)$$

Equation (5) indicates that the geometric phase holds for incident LP light with arbitrary

$\gamma$, providing a universality not achievable with resonant or propagation phases for linear polarization.

By contrast, the co-polarized component exhibits the same phase modulation only when $A = -B$, in which case the amplitude reduces to $Ae^{i2\sigma\tilde{\theta}}$. In practice, however, realizing an EP while simultaneously enforcing $A = -B$ is extremely difficult, rendering the co-polarized channel impractical for geometric phase modulation.

To realize EPs, we adopted the meta-atom design shown in Fig. 1e. The reflective metasurface follows a classical metal–insulator–metal (MIM) configuration, consisting of an aluminum ground plane, a silica spacer, and an array of aluminum nanoantennas. At an RCP (LCP) EP, polarization conversion from RCP to LCP (from LCP to RCP) is forbidden (Supplementary Note 2). Guided by this criterion, we systematically mapped the structural parameter space of the meta-atom to locate EPs. As shown in Fig. 2a, within the ($l_1$, $l_2$) parameter plane we identified a point where, for incident RCP light at 650 nm, the product of reflectance $r$ and polarization conversion efficiency (PCE)—defined as the ratio of the intensity of the reflected CP component with opposite handedness and the total reflected intensity—becomes zero. Simultaneously, the topological phase winds by $2\pi$ around this point, signifying the presence of an EP associated with degenerate RCP eigenstates. To further corroborate this, we computed the wavelength dependence of the eigenstates (Fig. 2b) and eigenvalues (Fig. 2c). At 650 nm, the two eigenstates coalesce into RCP with coincident complex eigenvalues, unambiguously confirming the existence of an EP.

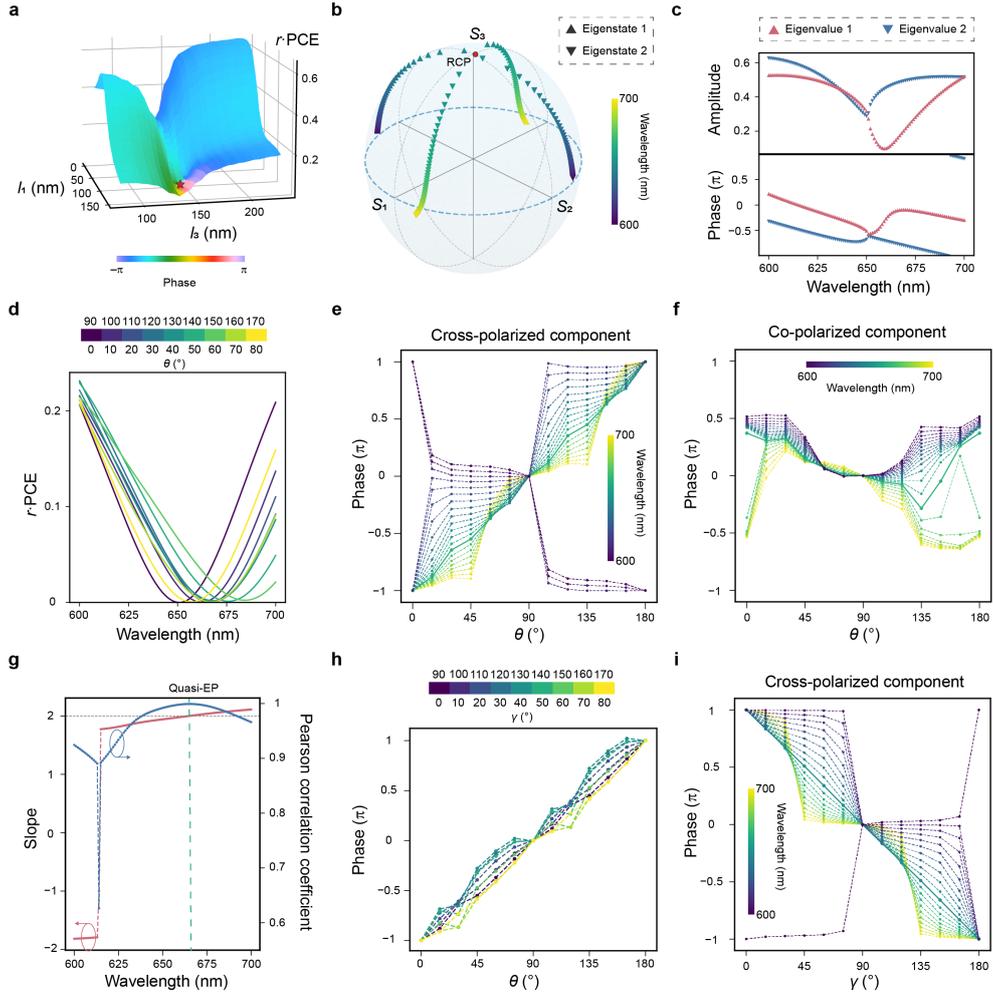

**Fig. 2 | Numerical simulation results for quasi-EP GPMs. a,** Simulated intensity (product of reflectance $r$ and PCE) and phase shift of the LCP output under RCP incidence, plotted as functions of $l_1$ and $l_3$ (other parameters as in Fig. 1e). An RCP EP (red star) occurs at $l_1 = 100$ nm and $l_3 = 145$ nm, where polarization conversion goes to zero and an encircling topologically protected $2\pi$ phase winding emerges. **b, c,** Simulated eigenstates (**b**) and eigenvalues (**c**) of the EP meta-atom as functions of wavelength. At 650 nm, the two eigenstates coalesce into RCP, and the complex eigenvalues coincide in magnitude and phase, confirming the EP. The ellipticity and azimuth of the eigenstates are given in Supplementary Note 3, further demonstrating their coalescence into RCP at 650 nm. **d,** Simulated spectra of $r \times$ PCE for meta-atoms with different rotation angles $\theta$. **e, f,** Phase shift of the cross-polarized (**e**) and co-polarized (**f**) outputs as a function of $\theta$ for $x$-polarized incident

LP light at different wavelengths. The quasi-EP wavelength (665 nm) is shown as a solid line, while other wavelengths are shown as dashed lines. **g,** Slopes of the phase–$\theta$ curves in (**e**) and the corresponding Pearson correlation coefficients as functions of wavelength. At the quasi-EP, the slope reaches 2 and the correlation coefficient attains its maximum, confirming accurate geometric phase control. **h,** Phase shift of the cross-polarized output as a function of $\theta$ for incident LP light at 665 nm with different polarization orientations $\gamma$, demonstrating that the quasi-EP GPM response is nearly independent of the input polarization angle. **i,** Phase shift of the cross-polarized output as a function of $\gamma$ at different wavelengths for the EP meta-atom with zero rotation angle. The response at the EP wavelength (650 nm) is shown as a solid line, with other wavelengths represented by dashed lines.

Establishing geometric phase modulation requires considering rotated meta-atoms. As shown in Fig. 2d, the zero of RCP-to-LCP conversion redshifts when the meta-atom is rotated relative to the fixed square lattice, indicating that the EP shifts with orientation. Consequently, a metasurface composed of differently oriented meta-atoms cannot maintain an exact EP but instead operates in a quasi-EP regime. We designate such metasurfaces as quasi-EP GPMs.

The quasi-EP-induced deviations of geometric phase modulation from Eq. (5) are minor and can be neglected for practical applications. Figs. 2e and 2g show that, for the cross-polarized output under $x$-polarized incidence, the phase–$\theta$ relation follows Eq. (5) most closely at 665 nm, both in slope and linearity. We identify this wavelength as the quasi-EP, redshifted from the EP at 650 nm of the unrotated meta-atom due to the EP shift illustrated in Fig. 2d. Away from the quasi-EP, deviations from Eq. (5) increase with spectral detuning. In contrast, Fig. 2f demonstrates that the co-polarized output does not yield practical phase modulation, in agreement with theoretical predictions.

Notably, Fig. 2h shows that quasi-EP GPMs continue to be effective for arbitrary LP incidence: although slight polarization-dependent deviations arise from shifts of the eigen-polarizations of the rotated meta-atom away from exact RCP, the phase modulation at 665 nm remains consistent with Eq. (5). This robustness establishes quasi-EP GPMs as a universal platform with respect to polarization angle for reliable geometric phase control of LP light.

Beyond this primary focus, Eq. (5) reveals an alternative route: for a homogeneous metasurface composed of EP meta-atoms with identical orientation, a geometric phase can be realized in the cross-polarized output simply by varying the polarization angle of the incident LP light. As shown in Fig. 2i, for the RCP EP meta-atom with zero rotation angle, varying the incident polarization produces an exact geometric phase at the EP wavelength. Full numerical simulations and analysis are provided in Supplementary Note 4. In this work, we concentrate—analogous to conventional GPMs—on geometric phases induced by rotating the meta-atoms. Nevertheless, this externally driven mechanism may open intriguing possibilities for future applications.

**Experimental validation of metagratings**

To experimentally validate the quasi-EP GPM concept, we fabricated a 321.6 μm × 321.6 μm metagrating with adjacent meta-atoms rotated by 30°, producing a phase gradient that deflects the beam to the +1$^{st}$ diffraction order (Fig. 3a). For direct comparison, a conventional GPM grating was designed under identical conditions (Supplementary Note 5). Scanning electron micrographs (SEMs) of the two devices are shown in Figs. 3b and 3d, and the corresponding optical microscope images are presented in Supplementary Note 6.

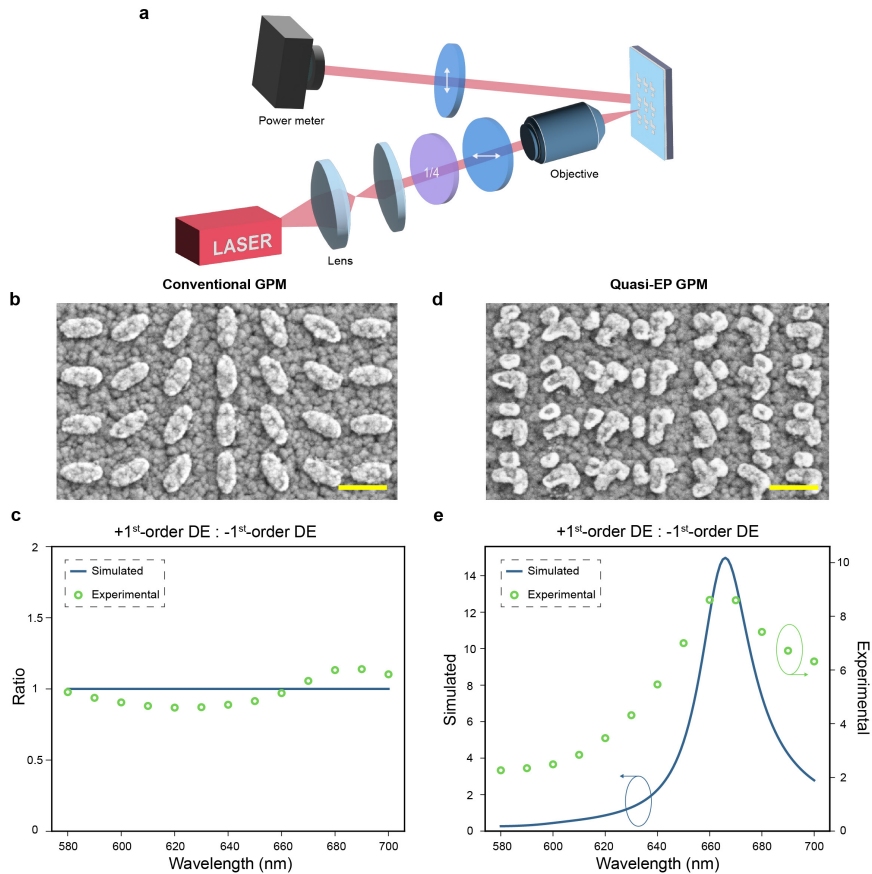

**Fig. 3 | Experimental validation of quasi-EP GPM grating. a,** Schematic of the optical setup for metagrating characterization, with the corresponding experimental setup photograph provided in Supplementary Note 7. **b,** SEM of a conventional GPM grating. Scale bar, 400 nm. **c,** Simulated and measured DE ratio between the +1st and -1st diffraction orders as a function of wavelength for the conventional GPM grating under the *x*-polarized LP incidence. Small discrepancies arise from fabrication imperfections and measurement uncertainties. **d,** SEM of a quasi-EP GPM grating. Scale bar, 400 nm. **e,** Simulated and measured DE ratio between the +1st and –1st diffraction orders as a

function of wavelength for the quasi-EP GPM grating under *x*-polarized LP incidence. The ratio peaks near the quasi-EP and stays above 3 across the spectrum, thereby validating the proposed concept.

Under *x*-polarized LP incidence, which decomposes into equal LCP and RCP components, the conventional GPM grating imposes opposite phase gradients on the two helicities (Eq. (1)), resulting in comparable intensities in the $+1^{st}$ and $-1^{st}$ diffraction orders (Fig. 3c). In contrast, the quasi-EP metagrating imparts a uniform geometric phase to the cross-polarized LP output, ideally steering all energy into the $+1^{st}$ order. Experimentally, the diffraction efficiency (DE) ratio of the $+1^{st}$ to $-1^{st}$ orders approaches 10 at the quasi-EP wavelength (Fig. 3e). Across the measured spectrum, the $+1^{st}$-order DE consistently exceeds the $-1^{st}$-order DE, confirming broadband operation. The maximum ratio is slightly lower than simulated values, and the spectrum around the quasi-EP is broadened—both of which are attributable to fabrication-induced deviations from the ideal EP condition. Notably, such broadening can benefit broadband applications. Furthermore, measurements under LP inputs with different polarization orientations $\gamma$ confirm universal operation with respect to $\gamma$, fully consistent with theoretical predictions (Supplementary Note 8).

**Experimental demonstration of metasurface holograms**

We next demonstrate quasi-EP GPM holograms operating in the Fraunhofer diffraction regime, thus extending the concept to complex wavefront shaping (Fig. 4a). For direct comparison, 320 μm × 320 μm quasi-EP and conventional GPM holograms were designed to reconstruct the letters "L" and "R", symmetric about the zeroth order. SEMs of the fabricated devices are shown in Figs. 4b and 4e, with the corresponding optical microscope images presented in Supplementary Note 6.

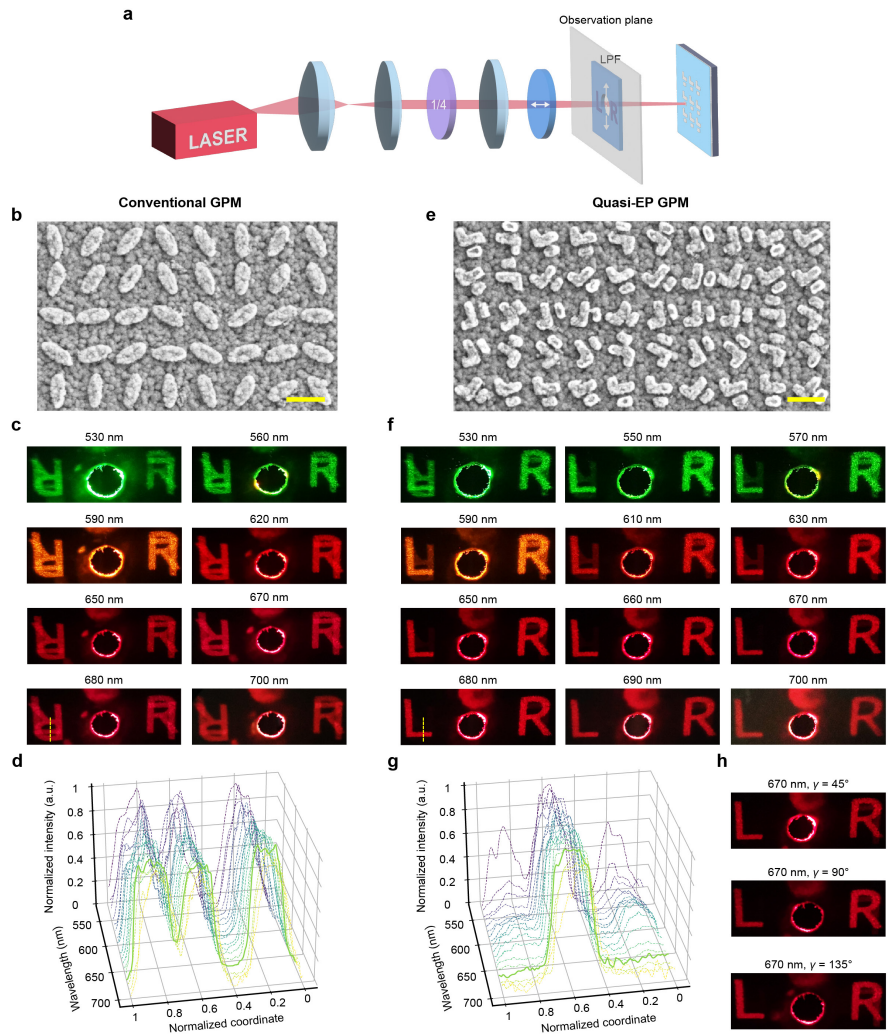

**Fig. 4 | Experimental demonstration of quasi-EP GPM hologram. a**, Schematic of the optical setup for metasurface hologram characterization, with the corresponding experimental setup photograph provided in Supplementary Note 7. LPF, linear polarizer film. **b,** SEM of a conventional GPM hologram. Scale bar, 400 nm. **c,** Experimental cross-polarized holographic images obtained from the conventional GPM hologram under *x*-polarized LP incidence across different wavelengths. Overlapping mirrored reconstructions lead to complete crosstalk across the spectrum. Complete

results are given in Supplementary Note 9. **d,** Intensity distributions along the line indicated in the lower-left holographic image in (**c**), for different wavelengths. The 670 nm curve is plotted as a solid line; other wavelengths are plotted as dashed lines (same convention in (**g**)). **e,** SEM of a quasi-EP GPM hologram. Scale bar, 400 nm. **f,** Experimental cross-polarized holographic images obtained from the quasi-EP GPM hologram under $x$-polarized LP incidence across different wavelengths. Crosstalk is negligible from 550 nm up to 700 nm, the upper limit of our measurement range (beyond which the signal-to-noise ratio was insufficient). Complete results are given in Supplementary Note 9. **g,** Intensity distributions along the line indicated in the lower-left holographic image in (**f**), confirming crosstalk-free, high-fidelity holographic display across a broadband spectrum. **h,** Experimental cross-polarized holographic images obtained from the quasi-EP GPM hologram at 670 nm under LP incidence with polarization angles of 45°, 90°, and 135°. Complete results are given in Supplementary Note 10.

Under $x$-polarized LP incidence, the conventional GPM hologram imparts opposite phase profiles to the decomposed LCP and RCP components. This produces diagonally mirrored holographic images that overlap completely, resulting in severe crosstalk across the spectrum (Figs. 4c and 4d). In contrast, the quasi-EP GPM hologram produces holographic images with negligible crosstalk in the cross-polarized LP output, sustained over a bandwidth exceeding 100 nm around the quasi-EP wavelength (Figs. 4f and 4g). When the operating wavelength deviates sufficiently from the quasi-EP wavelength, the quasi-EP GPM degrades toward conventional GPM behavior. This broadband performance does not conflict with the wavelength-dependent nature of phase modulation in Fig. 2e: the spatially varying distribution of meta-atoms breaks the periodic boundary condition, effectively broadening the quasi-EP bandwidth, while in practice most applications do not require strictly exact geometric phase control. Measurements with different LP orientations $\gamma$ further confirm that quasi-EP GPM holograms operate reliably for arbitrary linear polarizations (Fig. 4h).

We further tested the quasi-EP GPM hologram without polarization filtering at the output (Supplementary Note 11), corresponding to an ultracompact configuration that requires no external polarization optics when driven by LP sources. Although this increases crosstalk across the spectrum—highlighting the necessity of cross-polarized output selection—the image quality remains acceptable over a spectral window of several tens of nanometres around the quasi-EP wavelength.

## Discussion

In summary, we have introduced quasi-EP geometric phase metasurfaces that overcome the helicity constraint of the conventional PB phase and enable geometric phase control for linearly polarized light. Through simulations and experiments on metagratings and holograms, we confirmed high-fidelity wavefront shaping over a broad bandwidth and across diverse polarization orientations. In contrast to resonant and propagation phase metasurfaces for LP light, quasi-EP GPMs uniquely combine broadband functionality with polarization universality, establishing them as a practical solution for polarization manipulation and detection, imaging, holography, optical communications, and integrated photonic systems.

Beyond reflective devices, recent demonstrations of all-dielectric EP metasurfaces[30]

suggest that quasi-EP GPMs could be extended to transmission platforms and other architectures. In addition, our externally driven scheme for introducing geometric phases via polarization control (Supplementary Note 4) may open new application opportunities. Improving efficiency will require advanced design and optimization strategies, while device robustness remains constrained by the intrinsic sensitivity of EPs to fabrication errors. The central task of quasi-EP GPMs is to preserve polarization degeneracy at the EP while simultaneously enabling spatially varying geometric phase modulation. Interestingly, advances in bound states in the continuum (BIC) metasurfaces—demonstrating fabrication-robust ultra–high-$Q$ resonances[31] and arbitrary wavefront control under the quasi-BIC condition[32,33]—offer valuable inspiration, as the two fields, though seemingly unrelated, evolve along strikingly similar conceptual development paths. We anticipate that these concepts will accelerate the evolution of quasi-EP GPMs toward scalable, broadband, and practically deployable photonic platforms.

## Materials and methods

**Numerical simulations**

Simulations were performed using the finite-difference time-domain (FDTD) method. The Jones matrix of each meta-atom was constructed by extracting the reflected field components from two separate simulations with *x*- and *y*-polarized plane-wave excitation.

**Sample fabrication**

Polished silicon wafers were diced into 1 cm × 1 cm samples and cleaned. A 150-nm Al film was deposited via electron-beam evaporation, followed by a 50-nm $SiO_2$ spacer deposited using plasma-enhanced chemical vapor deposition (PECVD). After oxygen plasma treatment, a 180-nm PMMA A4 resist was spin-coated and patterned using 50-keV electron-beam lithography (EBL). The resist was developed in MIBK:IPA (1:3) for 1 min and rinsed in IPA. A 40-nm Al layer was then deposited via electron-beam evaporation. Lift-off was performed by immersing the sample in N-methyl-2-pyrrolidone (NMP) at 75 °C for ~1 h, followed by ~2 min of ultrasonic agitation.

**Optical characterization**

The optical system for characterizing the metagratings is illustrated in Fig. 3a, with a photograph of the experimental implementation provided in Supplementary Note 7. A supercontinuum laser (SC-Pro, YSL Photonics) combined with an acousto-optic tunable filter provided a wavelength-tunable beam, which was spatially filtered by an aperture, collimated by two lenses, and redirected by a mirror. A quarter-wave plate converted the incident linearly polarized light into circular polarization, and a linear polarizer produced linearly polarized light at arbitrary orientations. The beam was then focused onto the metasurface sample at an oblique angle using an objective lens. Diffracted beams of different orders were analyzed with a second polarizer and then detected with a photodetector.

The optical system for metasurface hologram characterization is shown in Fig. 4a, with corresponding setup photographs in Supplementary Note 7. The initial beam path was identical to that used for metagrating measurements, except that the collimated beam was normally incident on the sample and weakly focused onto it by a lens rather than an objective. The reflected holographic image was projected onto a screen with a central aperture that allowed the incident beam to pass through. A linear polarizer film with a matching central aperture was attached to the screen, leaving the polarization of the incidence unaffected while enabling polarization analysis of the reflected light.

## Acknowledgements

This work is supported by the National Natural Science Foundation of China (NSFC) grant [62222511], Natural Science Foundation of Zhejiang Province China grant [LR22F050006], National Key Research and Development Program of China grant (2023YFF0613000), and the STI 2030-Major Projects grant [2021ZD0200401]. The authors acknowledge support from the Micro-Nano Fabrication Center of Zhejiang University, in particular Jiabao Sun for assistance with thin-film deposition. The authors also thank Wei Wang for assistance with SEM measurements.

## Author contributions

Y.M. and Y.G. conceived the idea and designed the research project. Y.M. supervised the overall research. Y.G. carried out the theoretical analysis, simulations, and together with Q.C. fabricated the samples. Y.G. performed the experiments and analyzed the data. Y.M. and Y.G. initialized and edited the manuscript. All authors discussed the results and contributed to manuscript revision.

## Data availability

All key data supporting the findings of this study are included in the main article and its Supplementary Information. Additional datasets are available from the corresponding author upon reasonable request.

## Competing interests

Authors declare no competing interests.

Supplementary Information for

# Geometric phase metasurfaces for linearly polarized light

Yubin Gao[1], Qikai Chen[1], Yaoguang Ma[1,*]

## Affiliations

[1] State Key Laboratory for Extreme Photonics and Instrumentation, College of Optical Science and Engineering, ZJU–Hangzhou Global Scientific and Technological Innovation Center, Intelligent Optics and Photonics Research Center, Jiaxing Research Institute, Zhejiang University, Hangzhou 310027, China

[*] Corresponding author. Email: mayaoguang@zju.edu.cn

**The Supplementary Information include:**

Supplementary Note 1. Jones-matrix derivation of the geometric phase for circularly polarized light

Supplementary Note 2. Jones-matrix derivation of the geometric phase for linearly polarized light

Supplementary Note 3. Ellipticity and azimuth of simulated eigenstates of the EP meta-atom as functions of wavelength

Supplementary Note 4. External geometric phase by varying the polarization angle of the incident LP light

Supplementary Note 5. Design of conventional GPMs for comparison

Supplementary Note 6. Optical microscope images of metasurface samples

Supplementary Note 7. Experimental optical setups

Supplementary Note 8. Characterization of the quasi-EP GPM grating under LP incidences with different polarization angles

Supplementary Note 9. Complete characterization of the conventional and quasi-EP GPM holograms under $x$-polarized LP incidence

Supplementary Note 10. Characterization of the quasi-EP GPM hologram under LP incidences with different polarization angles

Supplementary Note 11. Characterization of the quasi-EP GPM hologram under $x$-polarized LP incidence without output polarization control

# Supplementary Note 1. Jones-matrix derivation of the geometric phase for circularly polarized light

We consider the configuration shown in Fig. 1a. In the linear polarization basis, the Jones vector of an incident CP wave can be written as

$$\vec{I} = \frac{1}{\sqrt{2}}\begin{bmatrix} 1 \\ i\sigma \end{bmatrix}, \tag{S1}$$

where $\sigma = \pm 1$ denotes LCP and RCP, respectively.

For a single-layer periodic meta-atom, the Jones matrix **J** can be expressed as

$$\mathbf{J} = \begin{bmatrix} A & C \\ C & B \end{bmatrix}, \tag{S2}$$

where $A$, $B$, and $C$ are generally complex coefficients. Owing to the out-of-plane mirror symmetry of the single-layer structure, the off-diagonal elements of **J** are equal.

When the meta-atom is rotated by an angle $\theta$, the output Jones vector is given by

$$\vec{O} = \mathbf{R}(\theta)\mathbf{J}\mathbf{R}(-\theta)\vec{I}, \tag{S3}$$

where $\mathbf{R}(\theta)$ is the rotation matrix, defined as

$$\mathbf{R}(\theta) = \begin{bmatrix} \cos\theta & -\sin\theta \\ \sin\theta & \cos\theta \end{bmatrix}. \tag{S4}$$

Substituting Eqs. (S1), (S2), and (S4) into Eq. (S3), we obtain

$$\vec{O} = \frac{1}{2\sqrt{2}}(A+B)\begin{bmatrix} 1 \\ i\sigma \end{bmatrix} + \frac{1}{2\sqrt{2}}(A-B+2i\sigma C)e^{i2\sigma\theta}\begin{bmatrix} 1 \\ -i\sigma \end{bmatrix}. \tag{S5}$$

We note that the polarization component with handedness opposite to the incident wave acquires an additional phase factor that explicitly depends on the rotation angle $\theta$. This phase corresponds to the conventional geometric phase for CP light:

$$\phi_{\text{conventional GPM}} = 2\sigma\theta. \tag{S6}$$

It is noteworthy from Eq. (S5) that when

$$A - B + 2i\sigma C = 0, \tag{S7}$$

the polarization component with handedness opposite to the incident wave vanishes, indicating that polarization conversion can no longer occur. The deeper physical implication of this condition will be discussed in Supplementary Note 2.

# Supplementary Note 2. Jones-matrix derivation of the geometric phase for linearly polarized light

The key to realizing geometric phase for LP light lies in controlling the eigen-polarizations of the meta-atom. We again consider the single-layer metasurface unit cell described by Eq. (S2). Let its eigenvalues and eigen-polarizations be denoted as $\lambda_{1,2}$ and $\vec{p}_{1,2}$, which satisfy

$$\mathbf{J}\vec{p}_{1,2} = \lambda_{1,2}\vec{p}_{1,2}. \tag{S8}$$

From standard linear algebra, the eigenvalues are obtained as

$$\lambda_{1,2} = \frac{1}{2}\left(A + B \pm \sqrt{(A-B)^2 + 4C^2}\right). \tag{S9}$$

An eigenvalue degeneracy ($\lambda_1 = \lambda_2$) occurs under the condition

$$(A-B)^2 + 4C^2 = (A - B - 2iC)(A - B + 2iC) = 0, \tag{S10}$$

which further gives

$$A - B - 2iC = 0, \tag{S11}$$

or

$$A - B + 2iC = 0. \tag{S12}$$

When $C = 0$, Eqs. (S11) and (S12) reduce to

$$A = B. \tag{S13}$$

In this case, the Jones matrix reduces to

$$\mathbf{J} = A\mathbf{I}, \tag{S14}$$

where $\mathbf{I}$ is the identity matrix. The eigenvalues are then $\lambda_{1,2} = A$, and any nonzero vector $\vec{p} \in \mathbb{C}^2$ satisfies $\mathbf{J}\vec{p} = A\vec{p}$. Thus, the entire $\mathbb{C}^2$ forms the eigenspace. This corresponds to a trivial degeneracy rather than an EP, and we therefore restrict our attention to the nontrivial case with $C \neq 0$.

For the first degeneracy condition expressed in Eq. (S11), the eigen-polarizations satisfy

$$\vec{p}_1 = \vec{p}_2 = \frac{1}{\sqrt{2}}\begin{bmatrix} 1 \\ -i \end{bmatrix}, \tag{S15}$$

corresponding to RCP.

For the second condition expressed in Eq. (S12), the eigen-polarizations are

$$\vec{p}_1 = \vec{p}_2 = \frac{1}{\sqrt{2}}\begin{bmatrix} 1 \\ i \end{bmatrix}, \tag{S16}$$

corresponding to LCP.

By redefining $\sigma = \pm 1$ for the cases where the degenerate eigenstate is LCP and RCP, respectively, the degeneracy condition can be compactly written as

$$A - B + 2i\sigma C = 0. \tag{S17}$$

Satisfying this condition gives rise to an EP in polarization whose degenerate eigenstate corresponds to the handedness $\sigma$. We note that this expression is identical to Eq. (S7). Physically, it implies that for a meta-atom supporting an LCP (RCP) EP, polarization conversion from LCP to RCP (from RCP to LCP) is prohibited. This provides a simple and reliable criterion for identifying EPs when scanning the parameter space of the unit-cell geometry (Fig. 2a).

With the meta-atoms discussed above, whose eigen-polarizations coalesce into a circular polarization state, it becomes possible to realize a geometric phase for LP light. Here we consider two equivalent scenarios. In the first, the metasurface is fixed while the incident and output linear polarizations rotate together by an angle $\gamma$. In the second, the polarizations are fixed while the periodic meta-atoms rotate by an angle $\theta$. In both cases, the input and output are orthogonal linear polarizations.

The Jones vector of the incident LP wave is

$$\vec{I} = \begin{bmatrix} \cos\gamma \\ \sin\gamma \end{bmatrix}, \tag{S18}$$

and the output Jones vector after interaction with the rotated meta-atom is

$$\vec{O} = \mathbf{R}(\theta)\mathbf{J}\mathbf{R}(-\theta)\vec{I}, \tag{S19}$$

where $\mathbf{J}$ satisfies Eq. (S17). Substituting Eqs. (S2), (S4), (S17), and (S18) into Eq. (S19) yields

$$\vec{O} = e^{i\sigma\tilde{\theta}}(A\cos\tilde{\theta} - i\sigma B\sin\tilde{\theta})\begin{bmatrix} \cos\gamma \\ \sin\gamma \end{bmatrix} + \frac{1}{2}i\sigma(A-B)e^{i2\sigma\tilde{\theta}}\begin{bmatrix} -\sin\gamma \\ \cos\gamma \end{bmatrix}, \tag{S20}$$

where we have introduced the relative rotation angle between the polarization reference frame and the metasurface orientation, $\tilde{\theta} = \gamma - \theta$, to simplify the expression.

From Eq. (S20), when the metasurface is rotated by a relative angle $\tilde{\theta}$, the orthogonal output component acquires a geometric phase given by

$$\phi_{\text{quasi-EP GPM}} = 2\sigma\tilde{\theta}. \tag{S21}$$

We refer to such metasurfaces as quasi-EP GPMs. In practice, rotation of the meta-atoms inevitably perturbs the ideal EP condition, meaning that the rotated structure only approximates an EP. As a result, the geometric-phase relation in Eq. (S21) holds approximately; nevertheless, the associated deviations are typically negligible for practical applications.

Equation (S20) also shows that for the output component parallel to the input polarization, geometric phase modulation of $2\sigma\tilde{\theta}$ occurs only under the condition $A = -B$, where the complex amplitude reduces to the form $Ae^{i2\sigma\tilde{\theta}}$. In contrast, when $A = B$, the coefficient reduces to $A$, corresponding to the trivial case of Eq. (S14), and no phase modulation occurs for either polarization component. In practical metasurface design and fabrication, simultaneously satisfying the EP condition while enforcing $A = -B$ is extremely difficult, making it challenging to exploit the co-polarized component for geometric phase modulation.

# Supplementary Note 3. Ellipticity and azimuth of simulated eigenstates of the EP meta-atom as functions of wavelength

For the RCP EP meta-atom (Fig. 1e), we calculated the ellipticity and azimuth of the simulated eigenstates shown in Fig. 2b, with the results plotted in Fig. S1. The ellipticity is defined as the ratio of the minor to the major axis of the polarization ellipse. At 650 nm, the ellipticity of both eigenstates approaches unity while the azimuth undergoes an abrupt change, indicating that the two eigenstates coalesce into RCP at this wavelength.

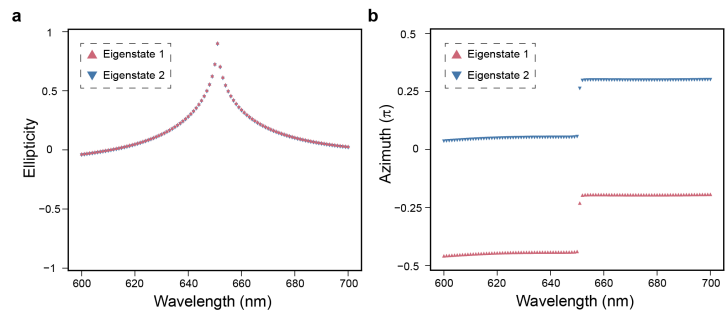

**Fig. S1 | Ellipticity (a) and azimuth (b) of the simulated eigenstates of the RCP EP meta-atom as functions of wavelength.**

# Supplementary Note 4. External geometric phase by varying the polarization angle of the incident LP light

From Eq. (S21), setting $\theta = 0$ yields

$$\phi_{\text{quasi-EP GPM}} = 2\sigma\gamma, \quad (S22)$$

which indicates that for a homogeneous metasurface composed of identically oriented EP meta-atoms, a geometric phase can be imparted to the cross-polarized output simply by varying the polarization angle $\gamma$ of the incident LP light.

To verify this prediction, we simulated a homogeneous metasurface composed of RCP EP meta-atoms (Fig. 1e) with zero rotation angle. Fig. S2 summarizes the results for incident LP light with varying polarization angle $\gamma$. As shown in Fig. S2a, the phase–$\gamma$ relation adheres most closely to Eq. (S22) at the EP wavelength 650 nm. Fig. S2b further shows that at 650 nm, the slope reaches −2 while the correlation coefficient attains its maximum, confirming accurate geometric phase modulation. By contrast, Fig. S2c demonstrates that the co-polarized output exhibits only negligible phase modulation, consistent with the theoretical expectations.

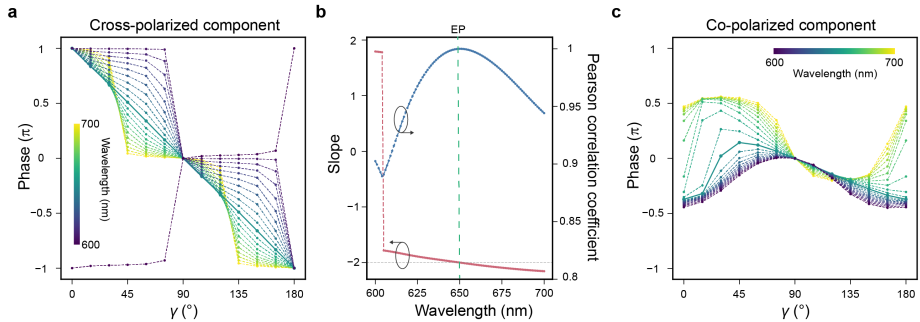

**Fig. S2 | Simulation results for the external geometric phase obtained by varying the polarization angle of the incident LP light. a,** Phase shift of the cross-polarized output as a

function of $\gamma$ at different wavelengths. The EP wavelength (650 nm) is shown as a solid line, while other wavelengths are indicated by dashed lines (the same convention is used in (**c**)). **b,** Slopes of the phase–$\gamma$ curves in (**a**) and the corresponding Pearson correlation coefficients as functions of wavelength. At EP wavelength, the slope reaches −2 and the correlation coefficient attains its maximum. **c,** Phase shift of the co-polarized output as a function of $\gamma$ at different wavelengths.

Unlike the quasi-EP GPM shown in Figs. 2e and 2g, no redshift of the quasi-EP occurs here. This is because only meta-atoms with zero rotation angle are employed, so the geometric phase relation is unaffected by the EP redshift in Fig. 2d that arises from rotating the meta-atom relative to the fixed lattice. In this case, external polarization rotation is equivalent to keeping the polarization unchanged while rotating both the meta-atom and the lattice together, so the EP condition remains unperturbed.

Taken together, these results confirm that varying the polarization angle of the incident LP light provides an external route to geometric phase control in EP metasurfaces, complementing the rotation-induced mechanism emphasized in the main text.

# Supplementary Note 5. Design of conventional GPMs for comparison

For comparison with quasi-EP GPMs, we designed a conventional GPM unit cell under the same conditions, as shown in Fig. S3a. The meta-atom adopts an elliptical pillar geometry. By scanning the structural parameters $l$ and $w$, we selected the design that maximizes both reflectance and PCE at the operating wavelength. The resulting reflectance and PCE spectra are shown in Fig. S3b.

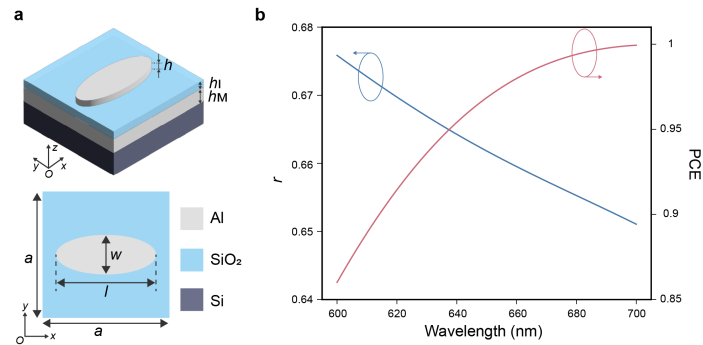

**Fig. S3 | Conventional GPM unit cell design. a,** Perspective (top) and top view (bottom) of the conventional GPM unit cell. Geometrical parameters: $h = 40$ nm, $h_\mathrm{I} = 50$ nm, $h_\mathrm{M} = 150$ nm, $a = 400$ nm, $l = 240$ nm, and $w = 120$ nm. **b,** Simulated reflectance and PCE spectra of the unit cell shown in (**a**).

# Supplementary Note 6. Optical microscope images of metasurface samples

Fig. S4 presents the optical microscope images of the metasurface samples shown in Fig. 3 and Fig. 4. In Fig. S4d, the quasi-EP GPM hologram exhibits a small amount of residual mask left after lift-off, but this has a negligible impact on the performance of the hologram.

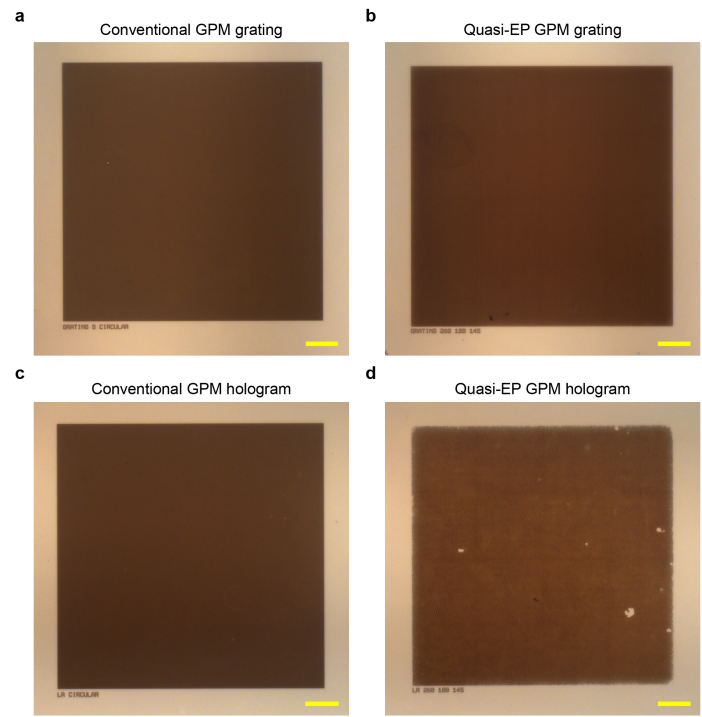

**Fig. S4 | Optical microscope images of metasurface samples.** Scale bar, 40 μm.

# Supplementary Note 7. Experimental optical setups

The optical setups for characterizing the metagrating (Fig. 3a) and the metasurface hologram (Fig. 4a) are shown in Figs. S5 and S6, respectively, with detailed descriptions provided in the Methods section.

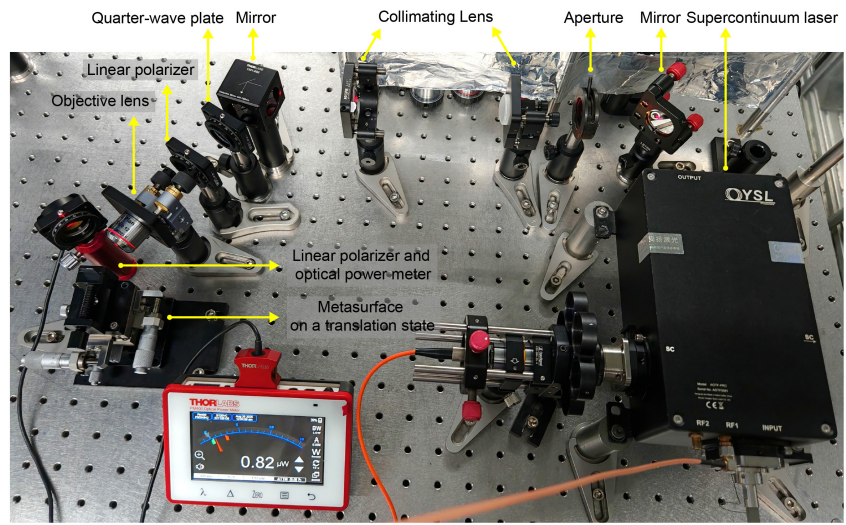

**Fig. S5 | Experimental optical setup for metagrating characterization.**

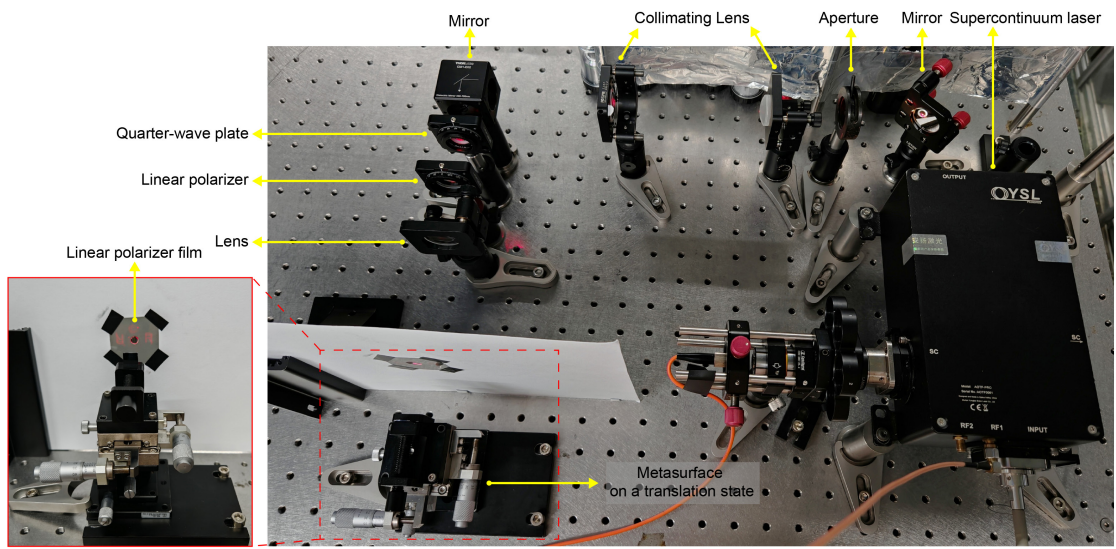

**Fig. S6 | Experimental optical setup for metasurface hologram characterization.**

# Supplementary Note 8. Characterization of the quasi-EP GPM grating under LP incidences with different polarization angles

To confirm that quasi-EP GPMs operate under arbitrary LP incidence, we characterized the quasi-EP metagrating using input beams with different polarization angles $\gamma$. The measured DE ratios of the cross-polarized output are plotted in Fig. S7. Compared with the $\gamma = 0°$ case in Fig. 3e, the spectral positions of the ratio maxima exhibit slight shifts, originating from deviations of the rotated meta-atom eigen-polarizations from exact RCP. Nevertheless, across the full spectral range the +1$^{st}$-order DE consistently exceeds the –1$^{st}$-order DE, verifying robust operation for multiple LP orientations. These experimental observations are fully consistent with the theoretical predictions, further confirming the linear-polarization-universal nature of quasi-EP GPMs.

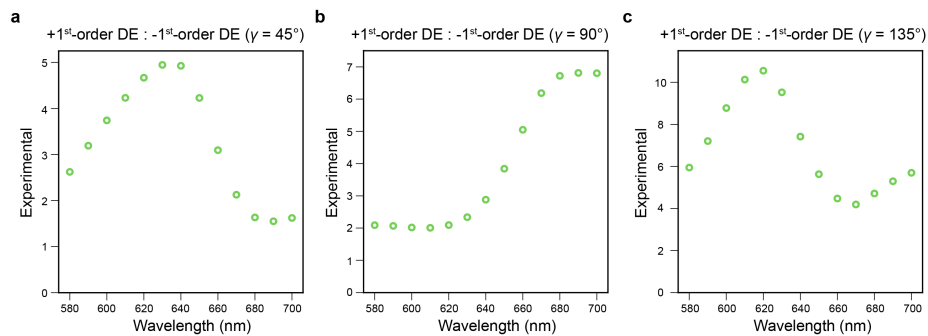

**Fig. S7 | Diffraction performance of the quasi-EP GPM grating under different LP incidences.** Measured DE ratio of the +1$^{st}$ and –1$^{st}$ diffraction orders as a function of wavelength for incident LP light with polarization angles $\gamma = 45°$ (**a**), $\gamma = 90°$ (**b**), and $\gamma = 135°$ (**c**).

# Supplementary Note 9. Complete characterization of the conventional and quasi-EP GPM holograms under *x*-polarized LP incidence

Figs. S8 and S9 present the complete characterization of the conventional and quasi-EP GPM holograms, respectively, under *x*-polarized LP incidence. Measurements were conducted from 530 nm to 700 nm in 10 nm increments.

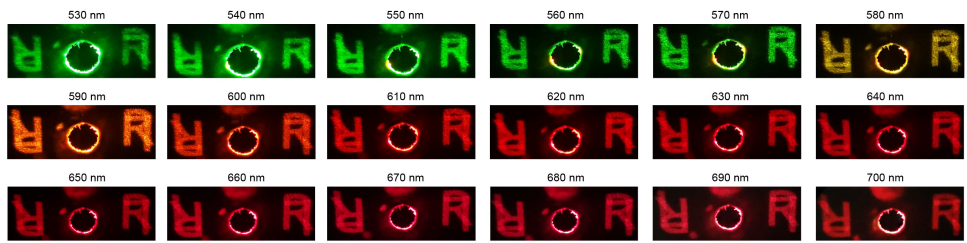

**Fig. S8 | Experimental cross-polarized holographic images obtained from the conventional GPM hologram.** Holographic reconstructions under *x*-polarized LP incidence at wavelengths from 530 nm to 700 nm in 10 nm steps.

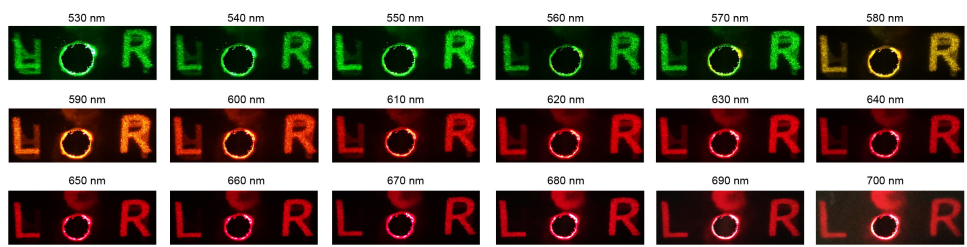

**Fig. S9 | Experimental cross-polarized holographic images obtained from the quasi-EP GPM hologram.** Holographic reconstructions under *x*-polarized LP incidence at wavelengths from 530 nm to 700 nm in 10 nm steps.

# Supplementary Note 10. Characterization of the quasi-EP GPM hologram under LP incidences with different polarization angles

To verify that quasi-EP GPM holograms operate under arbitrary LP incidence, we characterized the device under input beams with different polarization angles $\gamma$. Figs. S10–S12 present the cross-polarized holographic reconstructions at multiple wavelengths for incident LP light with $\gamma = 45°$, $\gamma = 90°$, and $\gamma = 135°$, respectively. These results demonstrate that the quasi-EP GPM hologram consistently produces broadband, high-quality holographic images for diverse linear polarization orientations.

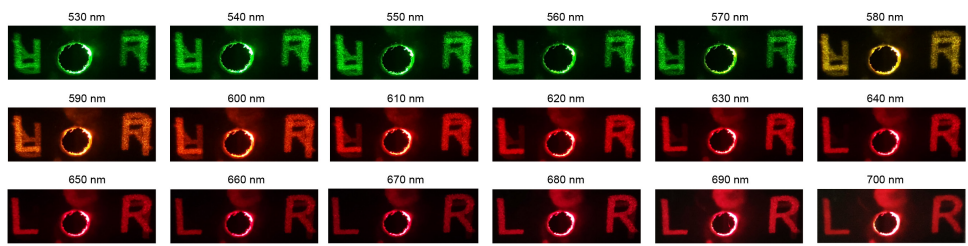

**Fig. S10 | Holographic performance of the quasi-EP GPM hologram under LP incidence at $\gamma = 45°$.** Experimental cross-polarized holographic images at different wavelengths.

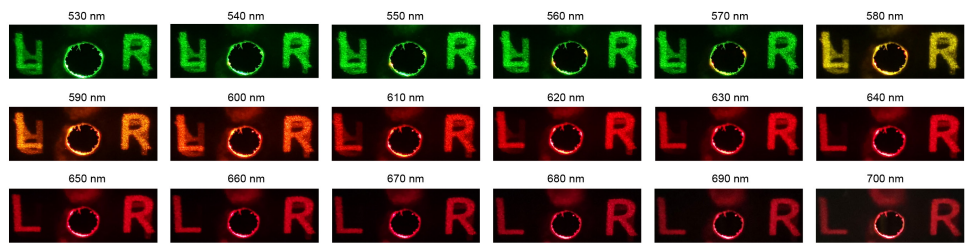

**Fig. S11 | Holographic performance of the quasi-EP GPM hologram under LP incidence at $\gamma = 90°$.** Experimental cross-polarized holographic images at different wavelengths.

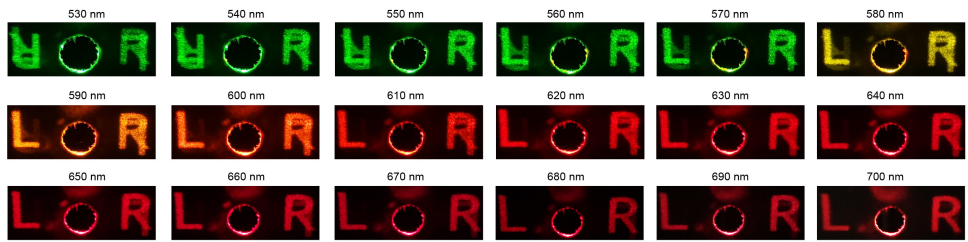

**Fig. S12 | Holographic performance of the quasi-EP GPM hologram under LP incidence at $\gamma = 135°$.** Experimental cross-polarized holographic images at different wavelengths.

# Supplementary Note 11. Characterization of the quasi-EP GPM hologram under *x*-polarized LP incidence without output polarization control

Fig. S13 shows the holographic images obtained from the quasi-EP GPM hologram without applying an output polarization filter. Compared with Fig. S9, where a linear polarizer film was used to select the cross-polarized component, the overlap of mirrored holographic images becomes more pronounced across the spectral range. Nevertheless, within a bandwidth of several tens of nanometres around the quasi-EP wavelength, the image quality remains acceptable. This feature is attractive for holographic applications with LP light sources, as it allows simplified optical configurations without external polarization optics.

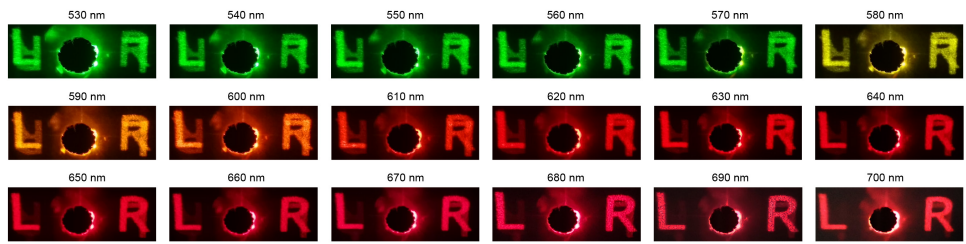

**Fig. S13 | Holographic performance of the quasi-EP GPM hologram without output polarization control.** Experimental holographic images at different wavelengths under *x*-polarized LP incidence, measured using the setup in Fig. S6 without the output linear polarizer film.